\documentclass[aps,prl,amsmath,amssymb,reprint,superscriptaddress]{revtex4-1}
\usepackage{graphicx}
\usepackage{amsmath}
\usepackage{dcolumn}
\usepackage{bm}
\usepackage{bbold}
\usepackage{color}
\usepackage{amsfonts}
\usepackage{amssymb}
\usepackage{mathrsfs}
\usepackage{tabularx}
\usepackage{braket}
\usepackage[T1]{fontenc}
  
\usepackage{caption}
\usepackage{subcaption}

\begin{document}

\title{Discovery of Iron Group Impurity Ion Spin States in Single Crystal Y$_2$SiO$_5$ with Strong Coupling to Whispering Gallery Photons}

\author{Maxim Goryachev}
\affiliation{ARC Centre of Excellence for Engineered Quantum Systems, University of Western Australia, 35 Stirling Highway, Crawley WA 6009, Australia}

\author{Warrick G. Farr}
\affiliation{ARC Centre of Excellence for Engineered Quantum Systems, University of Western Australia, 35 Stirling Highway, Crawley WA 6009, Australia}

\author{Natalia do Carmo Carvalho}
\affiliation{ARC Centre of Excellence for Engineered Quantum Systems, University of Western Australia, 35 Stirling Highway, Crawley WA 6009, Australia}

\author{Daniel L. Creedon}
\affiliation{ARC Centre of Excellence for Engineered Quantum Systems, University of Western Australia, 35 Stirling Highway, Crawley WA 6009, Australia}

\author{Jean-Michel Le Floch}
\affiliation{ARC Centre of Excellence for Engineered Quantum Systems, University of Western Australia, 35 Stirling Highway, Crawley WA 6009, Australia}

\author{Sebastian Probst}
\affiliation{Physikalisches Institut, Karlsruhe Institute of Technology, D-76128 Karlsruhe, Germany}

\author{Pavel Bushev}
\affiliation{Experimentalphysik, Universit\"{a}t des Saarlandes, D-66123 Saarbr\"{u}cken, Germany}

\author{Michael E. Tobar}
\email{michael.tobar@uwa.edu.au}
\affiliation{ARC Centre of Excellence for Engineered Quantum Systems, School of Physics, University of Western Australia, Crawley, 6009, Australia}

\date{\today}


\begin{abstract}

Interaction of Whispering Gallery Modes (WGM) with dilute spin ensembles in solids is an interesting paradigm of Hybrid Quantum Systems potentially beneficial for Quantum Signal Processing applications. Unexpected ion transitions are measured in single crystal Y$_2$SiO$_5$ using WGM spectroscopy with large Zero Field Splittings at 14.7GHz, 18.4GHz and 25.4GHz, which also feature considerable anisotropy of the \textrm{g}-tensors, as well as two inequivalent lattice sites, indicating spins from Iron Group Ion (IGI) impurities. The comparison of undoped and Rare-Earth doped crystals reveal that the IGIs are introduced during co-doping of Eu$^{3+}$ or Er$^{3+}$ with concentration at much lower levels of order 100 ppb. The strong coupling regime between an ensemble of IGI spins and WGM photons have been demonstrated at $18.4$~GHz and near zero field. This approach together with useful optical properties of these ions opens avenues for 'spins-in-solids' Quantum Electrodynamics.

\end{abstract}

\maketitle


The development of Hybrid Quantum Systems (HQS) has become a promising direction towards the realisation of the quantum information processing unit\cite{PhysRevLett.92.247902,PhysRevLett.102.083602,PhysRevLett.103.043603,Xiang:2013aa}. These systems usually require an optical or microwave electromagnetic coherent readout via 2D or 3D photonic cavities with severe requirements on system linewidths. Despite recent progress in superconducting structures, dielectric cavities supporting Whispering Gallery Modes (WGM) with Quality Factors of one hundred million ($10$~Hz linewidth) at the single photon level\cite{Creedon:2011wk} outperform the best superconducting 2D and 3D resonators limited by $Q$-factors of a few million\cite{Visser:2014aa,Megrant:2012aa,PhysRevLett.107.240501}. Due to such extremely low dissipation, WGMs are widely used as a probing tool in many areas of science including detection of mechanical motion\cite{Schliesser:2009aa}, { nanoparticle detection\cite{Li:2014aa,Ozdemir:2014aa} and sizing\cite{Zhu:2010aa}}, magnetic fields\cite{Forstner:2012aa}, biological substances\cite{He:2011aa,Baaske:2014aa}, test of fundamental physics\cite{PhysRevLett.95.040404} with sensitivities approaching the quantum limit\cite{Knittel:2013aa}, as well as classical\cite{Locke:2008aa,Nand:2013aa} and atomic\cite{BOURGEOIS:2006aa,Creedon:2010aa} oscillators. WGM resonators have been considered as microwave-to-optical up converters with one photon efficiency\cite{PhysRevA.80.033810}. 
Recently, these types of modes have been used for ultrasensitive microwave spectroscopy of paramagnetic impurities in dielectric crystals\cite{PhysRevB.88.224426,quartzG,PhysRevA.89.013810,Goryachev:2014aa,YAGfarr} with efficient sensitivity to detect naturally occurring impurities at the level of few parts per billion. This particular property make WGM system promising for HQS experiments involving long coherence photons and spins in solids. 

 Parameters of interaction between microwave photons and spins in dielectrics depends not only on the type of ion but also on the crystal host. 
 Single crystal Y$_2$SiO$_5$ (YSO) is a good candidate for this host role due for a number of reasons. Firstly, YSO is a low loss biaxial dielectric with a large enough dielectric constant (of order $10$)\cite{Natalia}. Such features of this crystal make it possible to design low loss 3D WGM type cavities in the X and K$_\text{u}$ frequency bands. Secondly, YSO crystal provide quiet spin environment for Rare Earth ions due to small magnetic moments of the constituent elements and small abundance of magnetic isotopes\cite{Bottger:2009aa}. Thirdly, Er$^{3+}$ ions in YSO crystal both have microwave (magnetic field controllable over the X and and K$_\text{u}$ bands) and infrared optical (telecommunication C-band) transitions exhibiting long coherence times\cite{PhysRevB.73.075101}. Fourthly, $^{167}$Er isotope has nonzero nuclear spin resulting in a hyperfine structure occupying 1-5 GHz range at zero external field\cite{Bushev:2011ve}. Due to combination of these microwave and optical properties, doped YSO crystals have recently drawn considerable attention in quantum optical community\cite{Zhong:2015aa,Clausen:2011aa,PhysRevLett.107.223202}. Potentially, they can be used for the physical realisation of microwave quantum memories\cite{Wolfowicz,Pr2015} and microwave-optical quantum interface\cite{PhysRevLett.113.063603,Williamson,PhysRevLett.92.247902,PhysRevLett.105.220501,PhysRevA.87.052333}. 

Despite the dominant role of Rare Earth dopings of YSO crystals for optical applications, the Iron Group Ions (IGIs) could also play an important role in some optical devices\cite{Kuo:1995aa}. In this work we discover a significant amount of unintentionally introduced spin impurities in Erbium and Europium doped single crystal (Er$^{3+}$:Y$_2$SiO$_5$, Eu$^{3+}$:Y$_2$SiO$_5$) which we attribute to IGIs (Chromium and Nickel ions in particular) due to existence of large Zero Field Splittings (ZFS). These ion impurities are unintentional co-dopants with the Rare-Earth Ions (REIs) introduced during the crystal growth process.
 


\begin{figure}[ht!]
	\centering
			\includegraphics[width=0.37\textwidth]{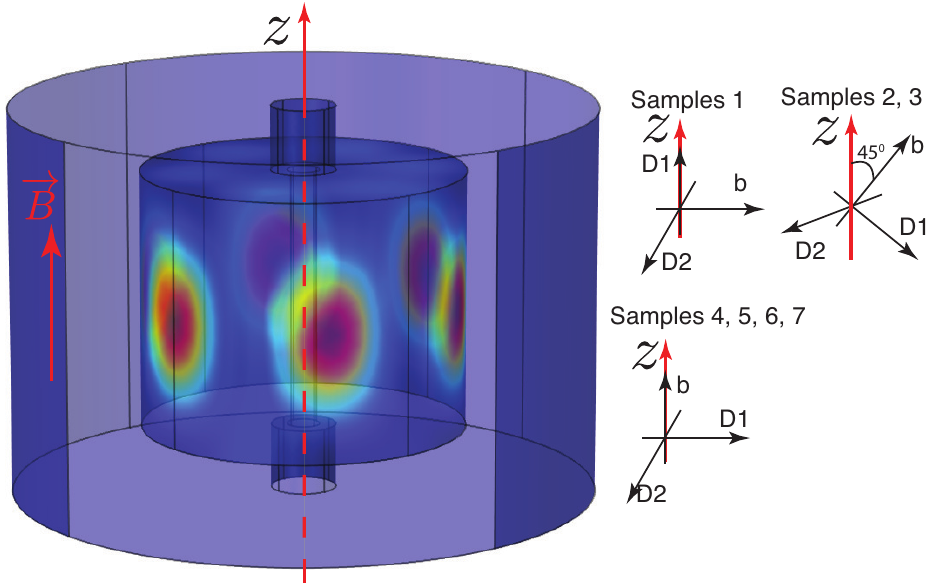}
	\caption{Simulation of a WGM in a YSO crystal inside a metallic cavity in terms of energy density, crystal axes orientation with respect to cylinder axes. }
	\label{expset}
\end{figure}

Experiments are performed with 7 cylindrically shaped YSO crystals (see Table~\ref{modesT} for details) grown by Scientific Materials Corp. The cylinders are placed in a copper shield situated in a superconducting magnet as shown Fig.~\ref{expset} and cooled to 20mK. More details of the experimental setup has been given previously\cite{Goryachev:2014aa,YAGfarr,PhysRevB.88.224426,quartzG}. 
 

The experimental procedure presuppose scanning of the external DC magnetic field and monitoring the cavity response\cite{PhysRevB.88.224426,quartzG}. When the splitting between energy levels approaches the WGM resonance frequency, the system exhibits an avoided crossing between two Harmonic Oscillators: one is the photon mode, the other is the spin ensemble\cite{ritsch}. If the coupling strength between a spin ensemble and a photonics mode exceeds the average linewidth of the resonances, the system is said to exhibit the strong coupling regime. This regime is characterised by hybridisation of the electromagnetic mode and the spin ensemble and is fruitful for many applications of quantum signal processing\cite{Wallraff:2004tw}.

\begin{figure}[ht!]
	\centering
			\includegraphics[width=0.5\textwidth]{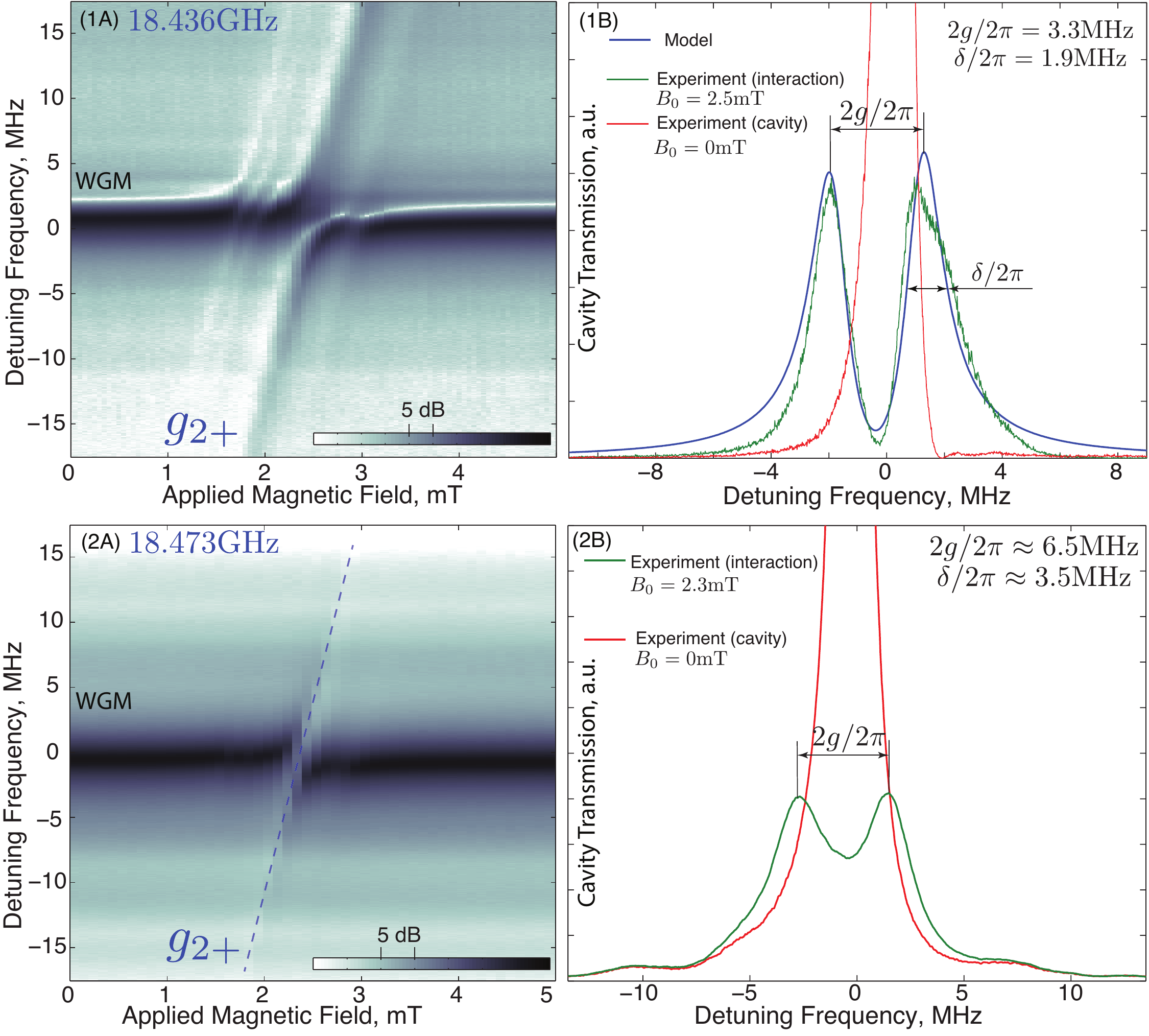}
	\caption{Interaction between WGM photons and IGI spins for (1) Sample 4 and (2) Sample 7 demonstrated as cavity transmission as a function of frequency and magnetic field (A) and as a function of detuning frequency at the interaction field (B).} 
	\label{strcoupl}
\end{figure}

\begin{figure*}[ht!]
	\centering
			\includegraphics[width=0.9\textwidth]{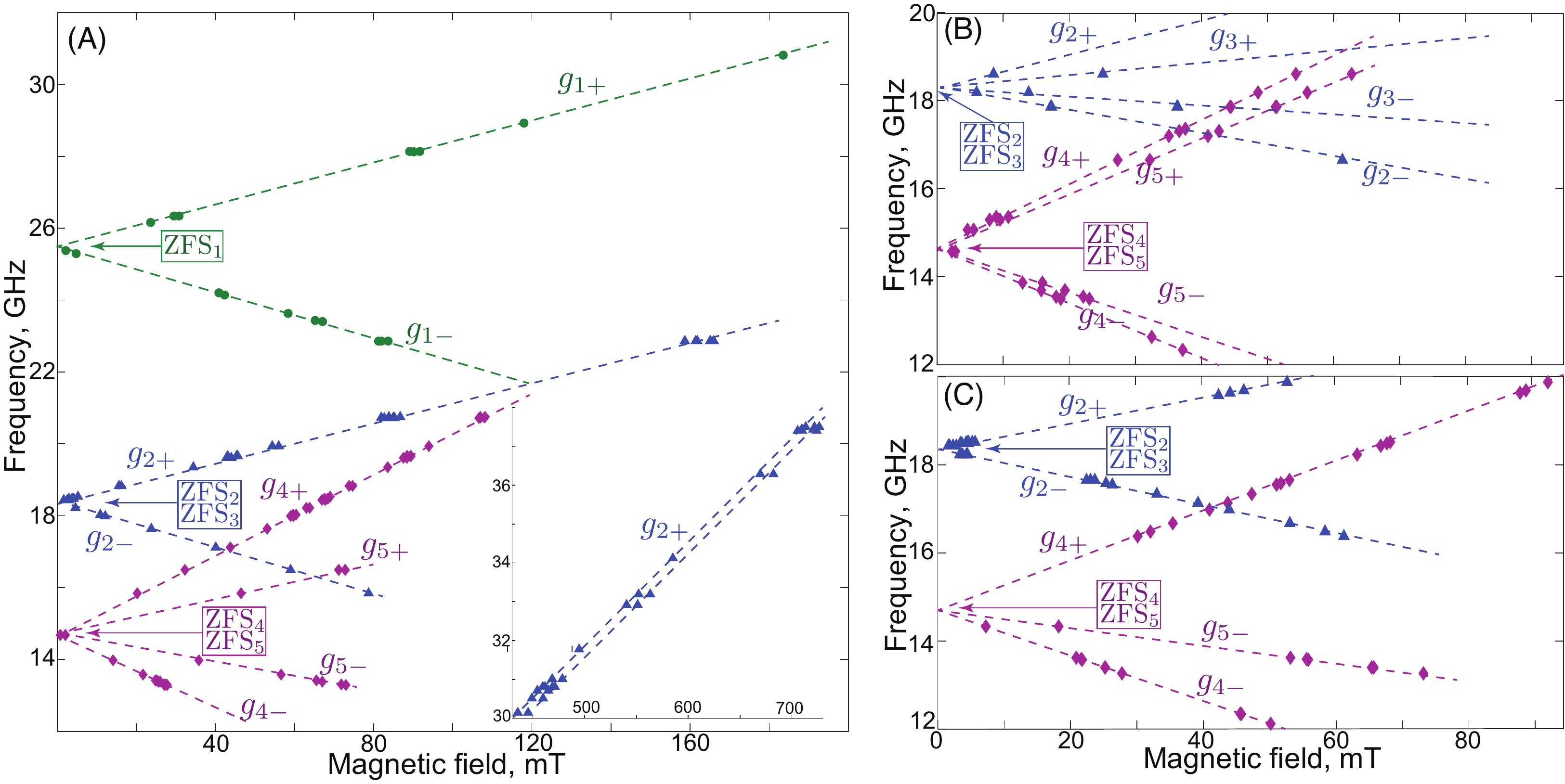}
	\caption{ALCs between IGI impurities in YSO crystals and WGM at 20 mK: (A) Sample 4 (The inset shows splitting of the $g_{2+}$ line), (B) Sample 2, (C) Sample 7. Er and Eu are not shown.}
	\label{spectra2}
\end{figure*}

{ The strong coupling regime is demonstrated in Fig.~\ref{strcoupl} (1A) and (1B), between a WG quasi-Transverse Magnetic (TM$_{1,2,1,-}$) mode at $\frac{\omega_0}{2\pi}=18.436$~GHz with a $Q$-factor of $3.7\times 10^4$, interacting with the $g_{2+}$ spins (see also Fig.~\ref{spectra2} for the detailed spectroscopy). Note, that due to the biaxial anisotropy of the crystal, the doublet pair due to the non-zero azimuthal mode number is non-degenerate (denoted as $\pm$)\cite{Natalia}. Here the TM$_{1,2,1,+}$ mode was measured to have a frequency of 18.528 GHz and exhibits a similar strong coupling but at 5.5 mT (not shown here)}. The corresponding normal mode splitting between the spin ensemble and the 18.436 GHz photon mode is $\frac{\text{g}}{\pi} \approx 3.3$~MHz which is greater than mean linewidth of the ensemble and WGMs ${2\delta} = \delta_{\text{WGM}}+\Gamma^{\ast}_2\approx 1.9$~MHz where standalone $\Gamma^{\ast}_2$ is $1.4$~MHz. {The corresponding average concentration of spins in the lower energy state can be estimated as $n=\frac{4\hbar}{\omega_0\mu_0\xi}\Big(\frac{g}{g_{DC}\beta}\Big)^2$,
where $\xi$ is the transverse magnetic filling factor, $g_{DC}$ is the DC g-factor, $\beta$ is the Bohr magnetron. The simulated transverse magnetic filling factor for both the quasi-TM$_{1,2,1,\pm}$ modes are about $0.5$, giving the corresponding concentration of the ions $4.5\times 10^{15}$~cm$^{-3}$ which is significantly less than the expected concentration of Er ions.
It should be noted that the external DC magnetic field at which the strong coupling regime is achieved is $2.5$mT. This value is lower than the critical field for the superconducting phase of Aluminium, and very favorable for the direct integration with aluminium SC quantum circuits\cite{PhysRevLett.107.220501,PhysRevLett.111.107008}. Observed photon $Q$-factors for both doped and undoped crystals is on the order of $10^5$ that is less than that for the state-of-the-art 2D and 3D superconducting resonators\cite{Visser:2014aa,Megrant:2012aa,PhysRevLett.107.240501}, but could be potentially improved in larger crystals with better filling factors. Such moderate $Q$-factors make it impossible to observe degradation of the cavity linewidths due to the doping.}

\begin{table*}[t]
\caption{Details of the crystal samples, ZFS, \textrm{g}-factors and spin linewidths $\Gamma^{\ast}_2$. Crystal dimensions are given in by radius $R$ and height $h$ and serial number by S/N. ND - Not Detected}
\centering
\begin{tabularx}{2\columnwidth}{X||X||X|X||X|X|X|X}
\hline
Sample &  1  &  2  &  3 & 4 & 5 & 6 & 7\\
\hline\hline
Orientation  &  D$1||z$  &  $\widehat{z\text{b}}=45^\circ$ & $\widehat{z\text{b}}=45^\circ$ &  b$||z$ & b$||z$ & b$||z$ & b$||z$\\
Doping, \%   &  Er$^{3+}$:0.005  &  Er$^{3+}$:0.001 & Er$^{3+}$:0.005 &  Er$^{3+}$:0.001 & Er$^{3+}$:0.001 & Er$^{3+}$:0.005 & Eu$^{3+}$:0.01\\
$R\times h$, mm   & $6\times 10$   & $5\times 10$ & $6\times 10$ &  $5\times 10$ & $5\times 10$  & $5 \times 7.67$ & $5\times 10$ \\
S/N   & 2-736-19   & 10-117-03 &  2-736-18 & 1-554-11  & 1-554-13  & 6-249-09 & 2-427-06 \\\hline
$\Gamma^{\ast}_2/2\pi$, MHz  &    &  & 1..2 &  1.4 & 1.1  &  & 2.2 \\
\hline
ZFS$_1$, GHz  &   &   &   &  25.44 &  & & \\
g$_{1\pm}$  &   &   & &  2.10/-2.22  &  & & \\
\hline
ZFS$_2$, GHz  & 18.38 & 18.38  & 18.36 & 18.38 & 18.38  & 18.43 & 18.38 \\
g$_{2\pm}$   & 1.94/ND  &  1.87/-2.25 & 1.86/-1.97 &  2.00/-2.29 &  2.00/-2.21 & 1.95/ND & 2.04/-2.32 \\
ZFS$_3$, GHz &   &  18.36 & 18.38  &  &  & & \\
g$_{3\pm}$  &   &  0.74/ND & 0.63/-1.03 &  &  & & \\
\hline
ZFS$_4$, GHz  & 14.69  & 14.70  & 14.71 &  14.68  & 14.74  & 14.74 & 14.69\\
g$_{4\pm}$   & 5.19/-4.38  & 5.21/-4.77  & 5.06/-4.55 &  4.05/-3.71 & 4.02/-4.00 & 3.78/-3.77 & 4.04/-3.71 \\
ZFS$_5$, GHz  & 14.68  & 14.67  & 14.72 &  14.72  & 14.70  & 14.71 & 14.68 \\
g$_{5\pm}$  &  2.24/-2.24 &  4.58/-3.72 & 4.41/ND & 1.75/-1.40 & 1.74/-1.42 & 1.76/-1.51 & ND/-1.4\\
\hline
\end{tabularx}
\label{modesT}
\end{table*}

The result of the experimental procedure described above can be represented by a number of avoided level crossings (ALCs) placed on a map where each dot denotes a crossing (see Fig.~\ref{spectra2}). Such a map is possible due to a large number of WGMs in a dielectric cylinder and their relatively narrow linewidths and high filling factors\cite{PhysRevB.88.224426}. A map of ALCs attributed only to the IGIs for two Er:YSO (samples 4 and 2), and Eu:YSO (sample 7), are shown in Fig.~\ref{spectra2} (A), (B) and (C) respectively, with corresponding Zeeman line interpretations. While the numerical parameter estimations for all seven samples are given in Table~\ref{modesT}. Classification of these ALCs as those belonging to IGIs is apparent from the structure of the plotted transitions, which exhibit large Zero Field splittings due to the crystal field significantly affecting the 3d electrons. In contrast, REIs have shielded 4f electrons resulting in the absence of ZFS for isotopes with zero nuclear spin. These measurements have been compared to spectroscopy of an undoped and purified YSO crystal. The spectroscopy demonstrated no ALCs that can be associated with REIs or IGIs. This suggest that both types of ions are introduced during the crystal growth and doping process for both the Erbium or Europium doped crystals. Note that IGI co-doping has not been previously observed\cite{Bushev:2011ve,Probst:2013zg,PhysRevB.90.100404}.

Not all ALCs can be classified as the strong coupling interaction due to mode differences in filling factors, polarisation and quality factors. Moreover, it is observed that typical coupling at larger external magnetic field is weaker than at lower ones due to the merging of magnetically inequivalent lines.

Spectroscopy in Fig~\ref{strcoupl} reveals the existence of three zero field splittings and five Kramers doublets. ZFS$_1$ appears at $25.4$GHz and corresponds to ESR lines $g_{1+}$ and $g_{1-}$, ZFS$_{2,3}$ appear at $18.4$GHz and corresponds to lines $g_{2\pm}$ and $g_{3\pm}$, ZFS$_{4,5}$ is at $14.7$GHz and corresponds to ESR lines $g_{4\pm}$- and $g_{5\pm}$. Fig. 2 (A) contains spectroscopic data taken when the magnetic field is parallel to the crystal axis. Therefore, all magnetically inequivalent positions merge. Since, no other ZFS has been found in the frequency range up to $30$GHz and also we found no evidence of ALC between spin states (as has been observed for Fe$^{3+}$ in sapphire\cite{PhysRevB.88.224426}), we conclude that observed spectrum is due to the impurities with $S=1$ and $S=\frac{3}{2}$.
Such spin numbers can only be associated wit Cr, Ni and Fe ions. There is evidence that the Chromium substitutes both the Y (octahedral site) and Si (tetrahedral site) with $3+$ and $4+$ valence states respectively. In particular, there are some ESR studies of such materials, especially Cr$^{4+}$:YSO\cite{Rakhimov:1999aa,Rakhimov:2000aa}, although detailed ESR spectroscopy of Cr:YSO is missing. Cr$^{3+}$ has $S=\frac{3}{2}$ and Cr$^{4+}$ has $S=1$.  Ni$^{2+}$ possess $S=1$ and does exist in 4-fold coordination and could theoretically (and at low levels) substitute for Si$^{4+}$ ions accompanied by an Oxygen vacancy for charge balance. Fe$^{6+}$ atomic radius is well matched to the Si$^{4+}$ radius, and this ion exists in the 4-fold coordination, however its $S=\frac{1}{2}$. Fe$^{4+}$ could also be excluded based
on the fact\cite{AYA:AYAA12967,Jia:1991aa} that the corresponding atomic radius is much larger than the native Si$^{4+}$, and it exists only in
6-fold coordination, rather than the native 4-fold of the Si site. Fe$^{3+}$ is not reported to exist in the coordination of the two native Y$^{3+}$ sites, 7-fold and 9-fold coordination. Fe$^{2+}$ with $S=3$ would yield complex ESR spectrum consisting of 3 Kramers doublets, one singlet state and many ALCs between spin states\cite{PhysRevB.88.224426}. 

Summarising this information with regards to the observed ZFS (Table~\ref{modesT}), it can be concluded 1) ZFS$_1$ (25.4 GHz) belongs to $S=1$ system which is most probably the Cr$^{4+}$ ion; 2) ZFS$_{2}$, ZFS$_{3}$ (18.4 GHz) most likely belongs to Ni$^{2+}$, $S=1$ system;
3) ZFS$_{4}$, ZFS$_{5}$ (14.7 GHz) is expected to be Cr$^{3+}$ giving $S = 3/2$ structure.

Despite the fact that the \textrm{g}-tensor is almost symmetric for IGIs, the observed ALCs demonstrate considerable dependence of on the crystal orientation. Indeed, in the case of the $g_{5\pm}$ lines, the effective DC \textrm{g}-factor changes by around the factor of two when the crystal axes are rotated by 45$^\circ$ angle. Table~\ref{modesT} shows that this dependence is consistent for crystals with the same orientation and different doping level. Although almost symmetrical \textrm{g}-tensors are typical for IGIs, the literature provides a few examples of significant magnetic anisotropy of IGIs in solids\cite{greenbook}.

Another feature that is not typical for IGIs is the splitting of the interactions into two lines at high field as shown for $g_{2+}$ on the inset of Fig.~\ref{spectra2}. This splitting is related to existence of two inequivalent sites for the same type of impurity ion. The difference between two \textrm{g}-factors is $0.03$ that becomes only resolvable due to presence of the high field ALCs shown on the inset. The splitting could be explained by slight misalignment with respect to the axis. { The temperature dependence of the IGI coupling strength follows the expected paramagnet spin ensemble dependence\cite{Sandner:2012aa,Goryachev:2014aa,PhysRevB.90.075112}.}

In conclusion, WGM spectroscopy of Er$^{3+}$ and Eu$^{3+}$ doped YSO crystals reveal additional impurities which cannot be attributed to REIs. Due to sufficient number of these impurities ions, the ensemble yields a strong coupling to WGMs at small fields. The coupling strength approaches $3.3$~MHz overcoming typical spin linewidth of $1-2$~MHz, which exceeds decay rates of SC quantum circuits. A spectroscopic map demonstrates ZFS attributed to Nickel and Chromium estimated to be present at the level of about 100~ppb. The measured g-factors reveal a strong anisotropy of these ions which is typical for anisotropic crystals. { The large measured ZFSs favour easier integration with SC qubits, as only very small magnetic fields need to be used to achieve strong coupling like in the case of Nitrogen Vacancies in diamond\cite{Zhu:2014aa}.}

We thank L. Alegria and Z. Cole (Scientific Materials Corp.) for the valuable discussions concerning of spin impurity doping of YSO crystals. This work was supported by Australian Research Council grant CE110001013, a UWA Research Collaboration Award and by BMBF project QUIMP 01BQ1060.


%

\end{document}